\documentstyle[psfig]{aa}

\newcommand{\be}{\begin{equation}}
\newcommand{\ee}{\end{equation}}
\begin{document}

\title{Complex Absorption Hides a Type II QSO in IRAS 13197-1627}
\author{M. Dadina
  \inst{}
  \and 
  M. Cappi 
  \inst{}}

\institute{IASF-CNR Sezione di Bologna, via Gobetti 101, 40129 Bologna, Italy\\ \email{dadina@bo.iasf.cnr.it}
\email{cappi@bo.iasf.cnr.it}}

\date{Received date/ Accepted date}

\abstract{

We report on the analysis of a BeppoSAX observation of IRAS 13197-1627. 
The complexity of the broad band (0.3-100 keV) spectrum of 
the source prevents us from selecting from two possible 
best-fit models, i.e. one based on a partial covering and the other on 
a reflection dominated scenario. Whatever the best fit model, the data imply 
that: 1) the primary spectrum is heavily obscured by at least two absorbers 
with 
column densities of the order of 5$\times$10$^{23}$  cm$^{-2}$ and 
$\geq$2$\times$10$^{24}$ cm$^{-2}$; 2) the intrinsic 2-10 keV luminosity of
the source is at least $\sim$2$\times$10$^{44}$ erg s$^{-1}$ thus making 
IRAS 13197-1627 a type-II QSO (more precisely a type 1.8 QSO)
rather than a 
Seyfert 1.8 galaxy as previously 
classified. Furthermore, there is marginal evidence 
of an absorption line  at E $\sim$7.5 keV suggesting a possible outflow 
with {\it v}$\sim$0.1c similar to recent findings on other QSOs and Seyferts.

  }
\maketitle

\section{Introduction}

Much of the research on Active Galactic Nuclei (AGNs) in recent 
years emphasized the search for Type-II QSOs. Unified models of AGNs 
(e.g. Antonucci 1993) explain the observed differences between broad 
(Type-I) and narrow (Type-II) emission-line AGNs as obscuration 
and viewing angle effects rather than intrinsic physical differences.
Although rather successful in describing nearby Seyfert galaxies, 
unified models seem to fail when extended to their higher-luminosity 
``relatives'', the QSOs (hereinafter, we shall broadly define a 
Seyfert as having 
L$_{\rm (2-10 keV)}<$10$^{44}$erg s$^{-1}$ and a QSO if L$_{\rm (2-10 keV)}>$10$^{44}$erg s$^{-1}$).

On the basis of unified models, the existence of a large population of type-II 
QSOs, or broadly speaking of a large population of heavily-absorbed 
high-luminosity 
AGNs has to be expected. However, because of the intrinsic difficulties to find them, only a limited number of such objects has been identified.
Their rarity is somehow embarrassing  considering that they are  
supposed to contribute the most to the 
total emission of the cosmic X-ray background. 
This is now supported both 
observationally from recent Chandra and XMM-Newton deep-fields 
and theoretically from synthesis models of the cosmic X-ray background 
(see e.g. the recent review by Fabian 2003). 

In this paper we report results from the BeppoSAX narrow field instruments
(Boella et al. 1997a) 
observation of IRAS 13197-1627. According to these, we find that 
the source can be identified  as the brightest Type-II QSO known to date. 
Given the above 
argument, it is thus an interesting case-study to understand the detailed 
absorption properties of this type of objects.

\begin{center}
\begin{table}[!ht]
\begin{minipage}{85mm}
\tabcolsep=3.0mm
\caption{Basic parameters of IRAS 13197-1627. (1) de Robertis et al. 1988; 
(2) Aguero et al. 1994; 
(3) Theureau et al. 1998.
}
\footnotesize
\scriptsize
\begin{tabular}{l|c}
\hline\hline
 & \\
Hubble Type  &        S? (1)   \\
 & \\
Seyfert Type & 1.8  (2)         \\
& \\
Redshift     & 0.01654   (3)   \\
 & \\
r.a. (2000)  & 13  22  24.462       \\
 & \\
Dec. (2000)  & -16 43 42.91     \\
 & \\
m$_{\rm v}$ & 14.46  \\
 & \\
\hline
\hline
\end{tabular}
\end{minipage}
\end{table}
\end{center}

IRAS 13197-1627, also named MCG-3-34-64\footnote{The source has been often 
misidentified with MCG-3-34-63, 
an edge-on galaxy $\sim$2$\arcmin$ north-est of IRAS 13197-1627. We checked that the 
X-ray emission is indeed coming from MCG-3-34-64 while MCG-3-34-63, is not 
detected in 0.1-10 keV band.}, has been firstly classified as a Seyfert 2 
galaxy ( Osterbrock \& de Robertis 1985, de Robertis et al. 1988) and, 
afterward, as a type 1.8 Seyfert  by Aguero et al. (1994) and  Young et al. 
(1996). Its basic parameters are given in Table 1.

The morphological classification of its host galaxy is unclear. 
In optical images it is extended over $\sim$30''$\times$30'' with only a few 
indications, if any, of spiral structures (de Robertis et al. 1988). 
A broad component in the H$\alpha$ line has been detected in polarized light 
by Young et al. (1996).

IRAS 13197-1627 was first observed in X-rays by ASCA (Ueno 1995). 
The ASCA spectrum was extremely steep (the photon index was 
$\Gamma$$\sim$3.0$\pm$0.3), with a high absorption column density of 
N$_{\rm H}$$\sim$4$\times$10$^{23}$ cm$^{-2}$ and with a soft fainter component 
emerging at energies below $\sim$3 keV. ASCA also detected an FeK$\alpha$ 
line at E$\sim$6.4 keV with an equivalent width EW$\sim$350 eV (Ueno 1995).

The present BeppoSAX dataset has been previously analyzed by Risaliti (2002) 
in a statistical study of a sample of 20 bright type-II Seyfert galaxies. 
However, Risaliti (2002) used only a predefined set of 3 spectral models and 
did not attempt a detailed modeling of all the spectral complexities found in 
this particular source. The reduced $\chi$$^{2}$ of the best-fit reported by 
Risaliti (2002) was $\sim$1.3, much larger than those reported here.  
The present work thus is complementary to and expands Risaliti's results.

\section{Observation and data reduction}

BeppoSAX observed IRAS 13197-1627 on July 22, 1998. 
The LECS (Parmar et al. 1997), MECS (Boella et al. 1997b) and PDS 
(Frontera et al. 1997) exposure times were of 14 ks, 44 ks and 19 ks 
respectively. 

Data reduction has been performed using the standard mission specific software 
SAXDAS based on the FTOOLS package. 
\begin{figure}
\psfig{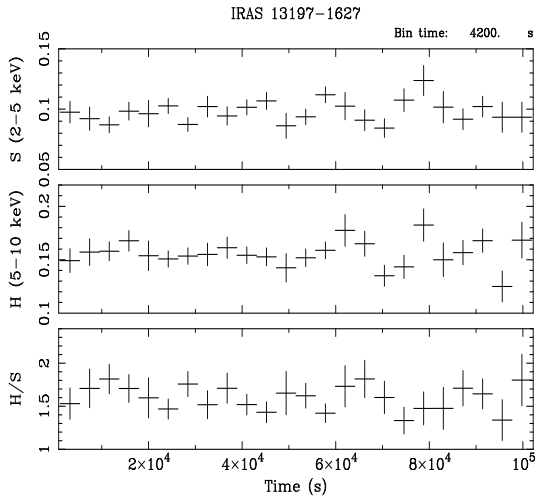}
\caption{MECS 2-5 keV (upper panel) 5-10 keV (middle panel) light curves and 
their ratio as a function of time (lower panel).}
\end{figure}

Two sources are visible at about 20$\arcmin$ from the target in the 
MECS 2-10 keV image. None is detected by the LECS 
so we can exclude that they contaminate the IRAS 13197-1627 counts 
at low energies. These sources are also not detected by the MECS in the 
5-10 keV energy range. 
For this reason they should not  contribute significantly  to the   
PDS counts. We checked in public archived databases that no other sources with 
known strong X-ray (2-10 keV) emission are present in the sky field covered by 
the PDS instrument (1.3$^{\circ}$ FWHM, Frontera et al. 1997).

Source counts have been extracted from circular regions centered on IRAS 
13197-1627. The LECS and MECS extraction regions have 
radii of 8$\arcmin$ and 4$\arcmin$ respectively. 
The LECS, MECS  and PDS spectra have been rebinned so as to sample
 the energy resolution of the instruments using grouping files produced by 
the ASI Science Data Center (ASDC) and taking care of having at least 20
counts for each obtained energy bin.
To subtract the background, we used standard PHA files accumulated from  
observations of empty sky regions and produced by the BeppoSAX team. We 
also checked that the results obtained using local backgrounds are  
consistent with the ones presented here. Most recent calibration files have 
been used for the spectral analysis.

\begin{center}
\begin{table*}
\hspace{3.0cm}\begin{minipage}{170mm}
\begin{center}
\footnotesize
\tabcolsep=3.0mm
\caption{SME model. (1) model name; (2) photon index; (3) absorption column 
density; (4) energy  of the absorption edge;  (5) optical depth of the 
absorption edge; (6) $\chi$$^{2}$/degrees of freedom. }
\scriptsize
\begin{tabular}{lccccc}
\hline\hline
&&&&&\\
Model&$\Gamma$& N$_{\rm H}$& E$_{\rm edge}$&$\tau$$_{\rm edge}$&$\chi$$^{2}$/d.o.f.\\
&&&&&\\
&&10$^{22}$ cm$^{-2}$ &keV& & \\
&&&&&\\
(1)&(2)&(3)&(4)&(5)&(6)\\
&&&&&\\
\hline
&&&&&\\
SME &1.90$^{+0.17}_{-0.15}$&50.9$^{+6.3}_{-5.7}$ &7.00$^{+0.13}_{-0.13}$&1.00$^{+0.24}_{-0.12}$&99.1/92\\
&&&&&\\
\hline
\hline
\end{tabular}
\end{center}

\end{minipage}
\end{table*}
\end{center}

Fig. 1 shows the MECS 2-5 keV (upper panel) and 5-10 keV (middle panel) 
lightcurves. The source flux is found to be roughly constant during the first 
half of the observation, while it is significantly (at a confidence level 
$\sim$95\% ) variable by up to 40\% on timescales down to 5-10 ks during the 
second half. 
These variations are observed 
both in the 2-5 keV band where the source spectrum is strongly affected by 
absorption and in the 5-10 keV energy range where the source should be
seen almost directly. 

As implied by the hardness ratios shown in Fig. 1 (lower panel), we found 
no significant spectral variations (with a confidence level $\geq$99\%) during the observation. We thus performed a time averaged spectral analysis collecting events from the whole 
period. 

\begin{figure}
\psfig{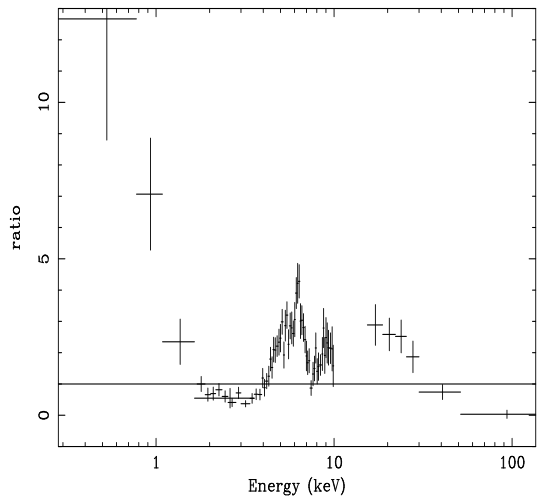}
\caption{Data-to-model ratios obtained when fitting the $\sim$0.3-100 keV data 
with a power-law model with Galactic absorption.}
\end{figure}

\section{Spectral analysis}

As shown by the data-to-model ratios plotted in Fig. 2, the broad-band 
($\sim$0.3-100 keV) spectrum of IRAS 13197-1627, once 
fitted with a simple power-law, clearly shows three main spectral 
features: 1) a soft excess below $\sim$2-3 keV; 2) a low-energy cut-off due to 
strong absorption between $\sim$3-5 keV; 3) a sharp and deep drop of the 
counts at E$\sim$7 keV.

\subsection{Simple Spectral Parametrization}

To account for the strong absorption and the soft excess we added to the simple
power-law  model a cold absorber (N$_{\rm H}$$\sim$5$\pm$1$\times$10$^{23}$ 
cm$^{-2}$) plus a power-law emerging at low energies. In a classical 
Seyfert 2 spectrum 
the latter component would be interpreted as due to scattering of the primary 
emission (hereinafter we will mention to this model as the 
``scattering model'' SM).
The data-to-model residuals (Fig. 3) for SM still show remaining 
complexities in the $\sim$4-8 keV energy range: an emission feature 
at 6.4 keV, a deep absorption edge at $\sim$7 keV and  
possible other components arising between $\sim$4-6 keV.

\begin{figure}
\psfig{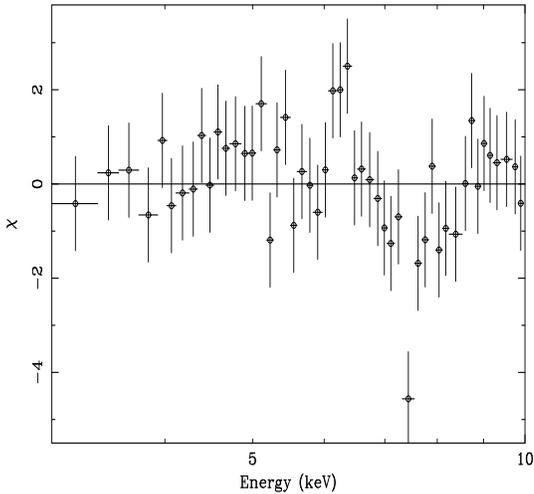}
\caption{The 3-10 keV residuals (expressed in terms of standard 
 deviation) for the SM. For clarity, data have been rebinned so as to 
have a 3$\sigma$ significance for each plotted data point.}
\end{figure}

To better model the data, we first tried to obtain a simple 
parametrization of all the features disregarding their physical 
implication.

The quality of the fit increases drastically after the addition of an 
absorption edge to the SM ($\Delta$$\chi$$^{2}$=24.3 for 2 additional
parameters, see SME model in Table 2). According to the F-test, 
this absorption edge is significant at $\geq$99.99\%  confidence level 
(see Fig. 4). The energy of the edge (E$\sim$7.0$\pm$0.2 keV) rules out 
(at $\geq$99\%) the presence of iron ionized at levels larger than FeX. 
If produced by the measured absorption column (N$_{\rm H}$$\sim$5$\times$10$^{23}$ 
cm$^{-2}$) the depth of the edge would require an iron overabundance ranging 
between 4 and 8 times the solar value.

\begin{figure}
\psfig{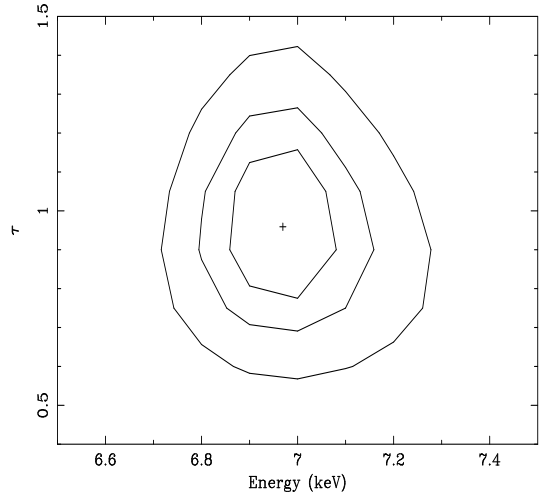}
\caption{Confidence contour levels (68, 90 and 99\%) for the edge optical depth $\tau$ versus its energy centroid.}
\end{figure}

\begin{figure}
\psfig{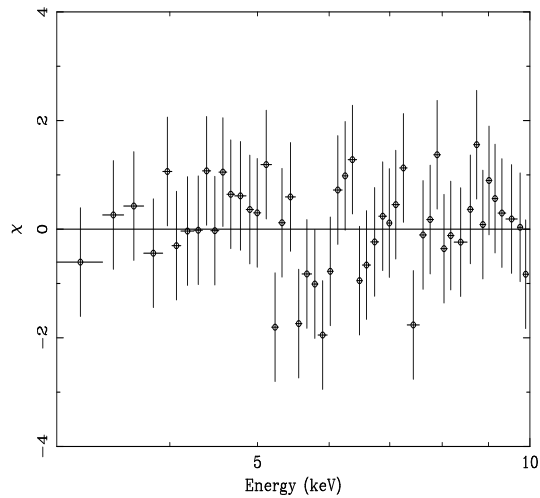}
\caption{Residuals in the 3-10 keV energy range for the SME model (data binned as in Fig. 3).}
\end{figure}

\begin{center}
\begin{table*}
\begin{minipage}{170mm}
\begin{center}
\footnotesize
\caption{Partial Covering Model. (1) model name; (2) photon index; (3) column density of the whole absorber; (4) covering fraction of the partial absorber; (5) column density of the partial absorber; (6) energy of the FeK$\alpha$ emission line; (7) equivalent width (EW) of the FeK$\alpha$ line; (8) $\chi$$^{2}$/degrees of freedom.}
\tabcolsep=3.0mm
\scriptsize
\begin{tabular}{l c  cccccc}
\hline\hline
&&&&&&\\

Model&$\Gamma$& N$_{\rm H,whole}$& C$_{\rm f}$& N$_{\rm H,partial}$&E$_{\rm FeK\alpha}$&EW&$\chi$$^{2}$/d.o.f.\\
&&&&&&\\
&&10$^{22}$ cm$^{-2}$ & &10$^{22}$ cm$^{-2}$ &keV&eV&\\
&&&&&&\\
(1)&(2)&(3)&(4)&(5)&(6)&(7)&(8)\\
&&&&&&\\
\hline
&&&&&&\\
PC&3.17$^{+0.34}_{-0.15}$& 47.8$^{+3.1}_{-2.1}$&0.94$^{+0.01}_{-0.03}$&269$^{+81}_{-55}$&6.42$^{+0.12}_{-0.12}$&228$^{+160}_{-179}$&82.9/90\\
&&&&&&\\
\hline
\hline
\end{tabular}
\end{center}
\end{minipage}
\end{table*}
\end{center}

Moreover, the data-to-model residuals for the SME model still show significant 
features between $\sim$4 and 8 keV (see Fig. 5). These features could be fitted 
by Gaussian emission lines (one broad at E$\sim$4.5 keV and one narrow at 
E$\sim$6.4 keV) and/or by Gaussian absorption lines (at E$\sim$6 keV and 
E$\sim$7.5 keV), all of which significant at more than 90\% confidence.

In summary, the ``simple spectral parametrization'' demonstrates that a single 
absorber cannot describe well the data above $\sim$2 keV.

\subsection{Partial Covering}

A possible explanation for the deep edge detected at $\sim$7 keV 
could be the occurrence of an 
additional heavy absorption that covers only partially the primary emission. 
This model fits well the data (model PC in Table 3 and Fig. 6) and the 
parameter values obtained indicate that 
about 95\% of the primary emission is absorbed by cold matter with 
N$_{\rm H}$$\sim$2$\times$10$^{24}$ cm$^{-2}$. The photon index of the continuum 
is quite steep ($\Gamma$$\sim$3.2$\pm$0.3) while the FeK$_{\rm \alpha}$ line 
is detected (at a confidence level $\geq$99\%) at E$\sim$6.4$\pm$0.1 keV 
with EW$\sim$228$\pm$180 eV. With this 
model the luminosity of the source is found to be 
L$_{\rm 2-10 keV}$$\sim$2$\times$10$^{44}$ erg s$^{-1}$. 
The observed FeK$_{\rm \alpha}$ line EW is lower
than expected for the observed column densities (an absorbing column 
of N$_{\rm H}$$\sim$2$\times$10$^{24}$ cm$^{-2}$  with a $\sim$95\% coverage
plus a N$_{\rm H}$$\sim$5$\times$10$^{23}$ cm$^{-2}$ with 100\% coverage should 
imply an EW$\sim$700 eV, Makishima et al. 1986).

\begin{figure}
\psfig{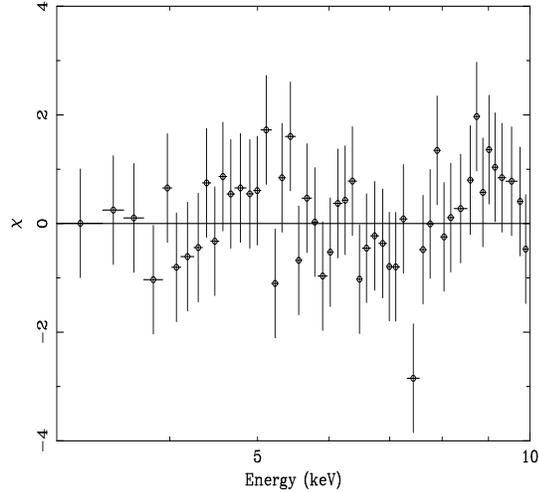}
\caption{Data-to-model residuals for the PC model (data binned as in Fig. 3).}
\end{figure}

\begin{table*}
\begin{minipage}{170mm}
\begin{center}
\caption{Pure Reflection model.  (1) model name;  (2) photon index;  (3) absorption column density;  (4) cut-off energy;  (5) energy of
the relativistic FeK$\alpha$ line;  (6) external radius of the region of emission of the relativistic line expressed in units of gravitational radii;  (7) equivalent width of the relativistic line;  (8) energy of the narrow FeK$\alpha$ line;  (9) equivalent width of the narrow line;  (10) $\chi$$^{2}$/degrees of freedom. $^{p}$ pegged parameter. $^{f}$ Fixed parameter.}
\footnotesize
\tabcolsep=3.0mm
\scriptsize
\begin{tabular}{lccccccccc}
\hline\hline
&&&&&&&&\\

Model&$\Gamma$& N$_{\rm H}$& E$_{\rm cut-off}$&E$_{\rm Laor}$&r$_{\rm out}$&EW$_{\rm Laor}$& E$_{\rm narr.}$&EW$_{\rm narr.}$&$\chi$$^{2}$/d.o.f.\\
&&&&&&&&\\
&&10$^{22}$ cm$^{-2}$ &keV&keV&keV&keV&keV&eV&\\
&&&&&&\\
(1)&(2)&(3)&(4)&(5)&(6)&(7)&(8)&(9)&(10)\\
&&&&&&\\
&&&&&&&&&\\
\hline
&&&&&&&&&\\
PR &2.30$^{+0.29}_{-0.31}$&41.2$^{+11.5}_{-10.8}$ & 76$^{+435,p}_{-42}$&6.40$^{f}$&6.05$^{+0.84}_{-0.83}$&3.81$^{+4.1}_{-1.9}$&6.40$^{+0.10}_{-0.10}$&356$^{+186}_{-143}$&66.8/88\\
&&&&&&&&&\\
\hline
\hline
\end{tabular}
\end{center}
\end{minipage}
\end{table*}

\subsection{Pure Reflection}

Alternatively to the PC model, the  edge at 7 keV could be 
produced by a  reflection component, which dominates the overall spectral 
shape. 
 
If we add to any of the above mentioned models a cold reflection component 
(namely PEXRAV model in XSPEC, Magdziarz \& Zdziarski 1995), we obtain 
that the $R$ parameter that define the relative 
importance of the reflected continuum with respect to the direct one, reaches 
very high values ($R$$\geq$10), thus reproducing a completely reflected 
spectrum.

\begin{figure}
\psfig{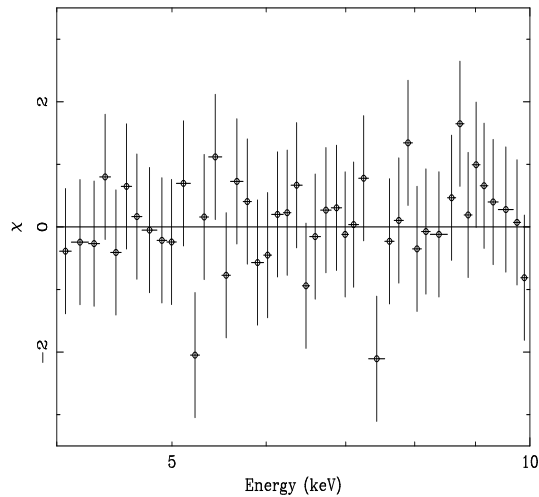}
\caption{Data-to-model residuals for the PR model (data binned 
as in Fig. 3).}
\end{figure}

We thus tried a model where the primary continuum is completely hidden 
for us and we only see the continuum reflected from the disk plus the 
associated FeK$\alpha$ 
line. This latter component has been parametrized using a relativistic model 
(the Laor model in XSPEC, see Fabian et al. 2002 for a recent review on this 
topic).  
Since the data clearly show a low energy cut-off, we also added an 
absorption column to the model plus the associated narrow Gaussian emission 
line.
To prevent the parameters of this model to diverge we fixed some values 
of the reflection component: metal abundance, R and inclination were fixed to
solar, 1 and 60$^{\circ}$ respectively. The only free parameters of the 
continuum  were thus the power-law photon index, the high energy cut-off and 
the normalization.
Similarly, we fixed many parameters of the relativistic line as follow: 
line energy E=6.4 keV, emissivity index i=-3, inclination angle=60$^{\circ}$, 
internal radius of the emitting region=1.25 gravitational radii r$_{g}$. The 
only free parameters of the Laor model thus were the external radius of the 
line emitting region (r$_{\rm out}$) and normalization.

This model well describes the data (PR model in Table 4 and Fig. 7) and the 
inferred intrinsic luminosity of the source is found to be 
L$_{\rm 2-10 keV}$$\geq$2$\times$10$^{44}$ erg s$^{-1}$. The fit results do not 
depend on the energy of the relativistic line. In fact, 
values of E$_{\rm Laor}$ from 6.4 to 6.9 keV give fits of the same 
quality ($\Delta$$\chi$$^{2}$$\leq0.2$).
The EW of the relativistic line is rather large  
(EW$\sim$3.8 keV) and it is in 
agreement with what is expected from a pure reflection spectrum (Matt et al. 1996). 
The outer radius of the line emitting region is found to be $\sim$5-7 r$_{g}$, 
a rather extreme value.
The energy and EW of the narrow line are in perfect agreement with what 
expected for a line produced via transmission in a cold absorber with the 
measured N$_{\rm H}$ of $\sim$4$\times$10$^{23}$ cm$^{-2}$  (Makishima 1986).

\subsection{An absorption line at 7.5 keV?}

In the data-to-model residuals both for the PC and PR model there is a marginal 
evidence of an additional absorption feature at E$\sim$7.5 keV. We checked 
this hypothesis by adding an absorption Gaussian line to the two models.

The quality of the fit drastically increases in the case of the partial 
covering ($\Delta$$\chi$$^{2}$=15.2 for 2 additional parameters, Fig. 8) 
while the feature seems to be marginal adopting the PR model as 
baseline ($\Delta$$\chi$$^{2}$=5.4). 

The feature is found to be at E$\sim$7.5$\pm$0.2 keV and to be narrow  
($\sigma$$<$ 0.47 keV) while its EW is $\sim$-330$\pm$180 eV
Its narrowness, in particular, seems to exclude that it could be associated to 
an unresolved additional edge due to warm matter. 

This absorption line could be associated with resonant absorption due to
He-like and/or H-like Fe. We tested this hypothesis modeling the feature
with the SIABS model (Kinkhabwala et al. 2003) assuming it is due to He-like
Fe. We obtained a good quality fit with physically plausible parameters, i.e. 
N$_{\rm Fe}$$\sim$10$^{19}$ cm$^{-2}$, outward velocity $v$$\sim$0.11c and 
dispersion $\Delta$$v$$\sim$4500 km s$^{-1}$.

\begin{figure}
\psfig{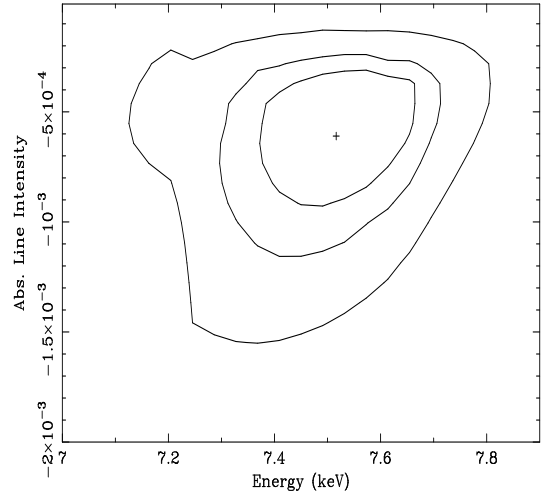}
\caption{Confidence contour levels for the absorption line at $\sim$7.5 keV in the PC model.}
\end{figure}

\section{Discussion}

The overall 0.3-100 keV spectrum of IRAS 1319-1627 is dominated by cold 
absorption and by an extra soft component that rises below $\sim$3 keV. 
This soft component is well modeled by a power-law with the 
same photon index of the hard component. Thus, as a first approximation, 
this spectrum appears as a very common (i.e. highly absorbed) 
X-ray spectrum of a Seyfert 2 galaxy where the low energy band is dominated by 
a scattered component (Smith and Done 1996, Turner et al. 1997, Risaliti 2002).

However, unlike other spectra of known Seyfert 2s, there is a major 
unexpected feature in the spectrum of IRAS 13197-1627 
that is a very deep absorption 
edge at $\sim$7 keV. The edge energy is consistent
with it being produced in a cold medium and rules out the presence of any
warm material. The depth of the edge requires an absorber with 
N$_{\rm H}$$\sim$2$\times$10$^{24}$ cm$^{-2}$, i.e. $\sim$3-4 times larger 
than what found by modeling
the low energy cut-off with a single absorption column.
Thus, to explain this edge, we are forced to 
invoke either that the source is partially covered or that it is completely 
obscured and its spectrum is dominated by the reflected component.

\subsection{The Pure Reflection and Partial Covering Scenarios} 

The pure reflection scenario seems rather extreme since a strong relativistic 
line (EW$\sim$ 3.8 keV) produced within $\sim$7 gravitational radii is 
required to fit the data.
Such an extreme line recalls the one that has been recently invoked for the 
narrow line Seyfert 1 MCG-6-30-15  (Wilms et al. 2001).

Bassani et al. (1999) suggested to use the ratio R$_{CT}$ between the 
absorption corrected OIII[$5007\AA$] flux and the observed 2-10 keV  flux
to identify Compton-thick sources. In their scheme, Compton-thick 
sources have R$_{CT}$$\leq$0.1. In the present case the 2-10 keV flux is 
$\sim$4$\times$10$^{-12}$ erg s$^{-1}$ cm$^{-2}$ and the absorption corrected 
F${[5077\AA]}$ flux is $\sim$4$\times$10$^{-12}$ erg cm$^{-2}$ s$^{-1}$ 
(Young et al. 1996) yielding R$_{CT}$$\sim$1. 
Also using the ASCA flux (Ueno 1995) one obtains R$_{CT}$$\sim$0.5.
Therefore, according to this criterium, IRAS 13197-1627 should not be
a Compton-thick source, making the pure 
reflection scenario even more implausible.

The partial covering scenario appears to be more likely 
not last for its simplicity. Nevertheless also in this case there are 
difficulties in interpreting the best fit parameters. First of all, the best 
fit photon index $\Gamma$$\sim$3 is rather steep for Seyfert galaxies 
(Smith and Done 1996, Turner et al. 1997) and it is only marginally consistent 
with what found for the most extreme Narrow Lines 
Seyfert 1 (NLSY1, Boller Brandt \& Fink 1996, Vaughan et al. 1999), i.e. the 
radio quiet AGNs with the known steeper spectra. 
It is worth noting here that another possible similitude with this class 
of sources is the observed variability ($\sim$40\%) down to
timescales of $\sim$5-10 ks. 
Another inconsistency in the partial covering scenario is that the EW of the 
FeK$\alpha$ 
line is lower than what expected for the observed absorption (see Sect. 3.2).

Present results show interesting analogies with what found 
for NLSY1s. This class of objects are generally characterized in X-rays
by a strong soft excess, a steep power-law and extreme variability 
(Boller et al. 1996, Brandt et al. 1997, Brandt \& Boller 1998, Vaughan et al. 1999). Recent
XMM-Newton results have added to this list, a new ``typical'' feature: a 
sharp spectral drop above 7 keV (Boller et al 2001, Pounds et al 2003a, Boller 
et al. 2003). 
Apart from the soft X-rays (that are here absorbed), 
IRAS 13197-1627 shows all the above mentioned 
characteristics of NLSY1s. The spectral complexity measured in IRAS 
13197-1627 is indeed very similar to what recently observed in MKN 766 (Pounds 
et al. 2003a), 1H 0707-495 (Boller et al. 2002) and IRAS 13224-38098 (Boller et 
al. 2003a) where  also in these cases the authors were left with two possible
scenarios that described the data equally well: either
 a reflection dominated or a partial covering scenario.

A theoretical explanation for the reflection dominated scenario has been 
recently proposed by Fabian et al. (2002). The main 
assumption of this model is that, in high accretion rate systems like NLSY1s, 
instabilities in the accretion may lead to the formation of 
multiple reflectors in the form of cold, dense material clumped 
into deep sheets or rings. 

Within the partial covering scenario, it could be that, as proposed by Pounds et al. (2003a) for MKN 766, dense ionized clouds or ejecta close to the 
super-massive black hole are 
responsible for the deep edge observed with XMM-Newton. Also in this case
our data are in good agreement with the model with the exception of the 
ionization state of the partially absorbing matter, our data requiring 
near-neutral Fe.

To disentangle between these two hypothesis better quality data above 3 keV
(e.g. from XMM-Newton) are needed.
Nevertheless, it is worth noting here that both scenarios imply 
that the intrinsic 2-10 keV luminosity of the source is 
$\geq2$$\times$10$^{44}$ erg s$^{-1}$. This makes IRAS 13197-1627 
the nearest and brightest type-1.8 QSO known to-date. This source thus
offers a unique possibility to study the spectral properties 
of this class of objects and to 
test unified models at the bright end of the 
luminosity function of Seyfert galaxies. 
This is of particular importance given that these objects 
are thought to contribute the most of the cosmic X-ray background. 
IRAS 13197-1627 is thus  
an important case study to understand what the absorption characteristics of
luminous QSO could be.

\subsection{Possible Origin of the Complex Absorption}

As stated above, the spectral analysis shows 
that the source has heavy and
complex absorption that cannot be modeled by a single absorber. 

We speculate that the most plausible scenario is the one in which 
the lower column is due to the putative dusty torus while the 
heavier absorption is due to dense clouds, blobs or winds close to the  X-ray 
emitting region. The presence of a
dusty torus, in fact, would explain the different columns measured in X-rays  
(N$_{\rm H}$$\sim$5$\times$10$^{23}$ cm$^{-2}$) and in optical (the 
E$_{\rm B-V}$ in the direction of the nucleus implies a  
N$_{\rm H}$$\sim$1.5$\times$10$^{21}$ cm$^{-2}$, Lumsden et al. 2001). On the 
other hand, dense blobs or clouds close to the X-ray source may either 
partially cover or block the nuclear emission without obscuring the 
component reflected by the accretion disk.  

It is worth noting here that this picture would also be able
to explain, if confirmed, the absorption feature at $\sim$7.5 keV. 
This, in fact, could be interpreted as He- or H-like Fe resonant absorption by 
outflowing matter. 
The inferred velocity of this outflow then should be $\sim$0.1c, 
similar to what found in PG1211+143 (Pounds et al. 2003).

\section{Conclusion}

The BeppoSAX X-ray spectrum of IRAS 13197-1627 is so complex that we could not 
find a single best-fit 
model and were left with two possible scenarios: {\it i}) one 
where the primary emission is completely blocked and the resulting spectrum is 
completely reflection dominated and {\it ii}) one
where the spectrum is due to a partial covering of the X-ray source.

Despite this ambiguity, a firm conclusion of this work is that any modeling of 
the X-ray spectrum 
implies an intrinsic luminosity  of at least
L$_{\rm 2-10keV}$$\sim$2.0$\times$10$^{44}$ erg s$^{-1}$ thus making IRAS 
13197-1627 the nearest and brightest example of a type-II QSO.

Moreover, both scenarios require a complex absorption structure composed of at 
least two heavy absorbers. We speculate that these could be a dusty  
torus and clouds, blobs or winds closer to the nucleus.
Finally, if  the absorption line marginally detected
at 7.5 keV is identified with  resonant absorption from He- or H-like Fe, 
then we infer that some of matter (e.g. the clouds or the wind) should have an 
outward velocity of the order of $\sim$0.1c.

\acknowledgements  
We are very grateful to G. G. C. Palumbo and L. Foschini for useful 
discussions. MD gratefully acknowledge the Italian Space Agency (ASI) for 
financial support under contract I/R/042/02. This research has made use of t
he ASDC (ASI Science Data Center) database.

\end{document}